\documentstyle [12pt,epsfig]{article} 
\makeatletter
\@addtoreset{equation}{section}
\makeatother
\renewcommand{\theequation}{\thesection.\arabic{equation}}
\newcommand{\BS}{\bigskip}
\newcommand{\SECTION}[1]{\BS{\large\section{\bf #1}}}

\begin{document}
\begin{titlepage}
\begin{center}
{\large \bf
Convergence and Gauge Dependence Properties of the
Resummed One-loop Quark-Quark Scattering
Amplitude in Perturbative QCD }
\vspace*{1.5cm}
\end{center}
\begin{center}
{\bf J.H. Field }
\end{center}
\begin{center}
{ 
D\'{e}partement de Physique Nucl\'{e}aire et Corpusculaire
 Universit\'{e} de Gen\`{e}ve. 24, quai Ernest-Ansermet
 CH-1211 Gen\`{e}ve 4.
}
\end{center}
\vspace*{2cm}
\begin{abstract}
The one-loop QCD effective charge $\alpha_s^{eff}$ for quark-quark scattering
is derived by diagrammatic resummation of the one-loop amplitude using an 
arbitary covariant gauge. Except for the particular choice of gauge parameter
$\xi = -3$, $\alpha_s^{eff}$ is found to {\it increase} with increasing
 physical scale, $Q$, 
as $\ln Q$ or $\ln^2 Q$. For $\xi = -3$, $\alpha_s^{eff}$  decreases with 
increasing $Q$ and satisfies a renormalisation group equation. Also, except for the case
 $\xi = 19/9$, convergence radii of geometric series are found to impose upper limits on $Q$.
\end{abstract}
\vspace*{1cm}
\newline
{\it Keywords ;} Quantum Chromodynamics
Renormalisation Group Invariance,
Asymptotic Freedom.
\end{titlepage}

\SECTION{\bf{Introduction}}
 Quark-quark scattering in next-to-leading order QCD has been calculated
 by several different groups~[1-4]. 
 Coqueraux and De Rafael~\cite{x1}, calculated the one-loop corrections to the
 invariant amplitude in the Feynman gauge using an on-shell renormalisation
 scheme~\cite{x5}. In Refs.~[2-4] complete expressions for the squared invariant
 amplitude were given in the dimensional regularisation scheme~\cite{x6}.
 The calculations presented in the present paper generalise those of Refs.~[1-4]
 in two ways:  
 \begin{itemize}
 \item[(i)] An arbitary covariant gauge is considered.
 \item[(ii)] The one-loop Ultra-Violet (UV) divergent loop and 
  vertex diagrams are resummed to all orders in $\alpha_s$.
  \end{itemize}
  (ii) yields a scale-dependent `effective charge' as a factor in the
  invariant amplitude. The resummation is done, not by solving a 
  renormalisation group equation (RGE), but by an exact sum of the relevant 
  diagrams to give the QCD analogue of the Dyson sum of QED.  
  \par The results are very surprising. The effective charge is 
  gauge dependent at $O(\alpha_s^3)$ and beyond, and except for one
  specific choice of gauge, does not display `asymptotic freedom' but
  instead {\it increases} as $\ln Q$ or $\ln^2 Q$ at large scales $Q$. 
  Only for the same special choice of gauge, where contributions from
  vertex diagrams vanish, does the effective charge satisfy a RGE of the
  type that is valid in 
  QED~[7-11]. For this special choice of gauge (called `loop gauge')
  the effective charge decreases with increasing scale, but only to 
  a fixed limit $Q_L$, determined by the convergence radius of the 
  geometric sum of gluon and fermion loops in the dressed gluon 
  propagator. The measured value of $\alpha_s$
  suggests that $Q_L \simeq$ 300 GeV, already of phenomenological 
  importance at the Fermilab $\rm{p} \overline{\rm{p}}$ collider.

  \par The plan of the paper is as follows. In the next Section the
   UV divergent (before renormalisation) one-loop corrections to the
   quark-quark scattering amplitude in Feynman gauge are derived from similar
   corrections to the quark-quark-gluon vertex given in Ref.~\cite{RDF}. The 
   generalisation to an arbitary covariant gauge is made using results
   reported in Ref.~\cite{x13}. In Section 3, the one-loop corrections
   are diagrammatically resummed to yield the QCD analogue of the Dyson sum
   of QED. In Section 4 the self-similarity and renormalisation group
   properties of the effective charge derived in Section 3 are discussed.
   In the final Section the classical proofs of the asymptotic
   freedom property of QCD in the literature are critically examined 
   in the light of the results obtained in the previous Sections. Also briefly
   discussed are : (i) `renormalons' (ii) the generalisation to higher
   loop order 
   vacuum polarisation and vertex corrections and (iii)
    pinch technique and related calculations of proper self energy and 
   vertex functions, both in QCD and in the Standard Electroweak Model.
    \par The present paper is the fifth in a series on fundamental physics aspects of perturbative QED and
    QCD: Ref.~\cite{JHFAP} discusses on-shell renormalisation and optimised perturbation theory
     in QED and QCD; Ref.~\cite{JHFIJMP1} considers the role of non-vanishing fermion or gluon
     masses as physical regulators for diagrams with combined infra-red (IR) and ultra-violet (UV)
     divergences; the relation of such IR/UV divergent diagrams to the Lee-Naunberg~\cite{LN} and
     Kinoshita~\cite{Kinoshita} Theorems of QED and QCD is examined in Ref.~\cite{JHFMPLA};
     convergence conditions for resummed physical amplitudes, as discussed in the present paper
     for quark-quark scattering in QCD, are analysed for the analogous QED case of scattering of
     unequal mass charged fermions in Ref.~\cite{x12}.
    
\SECTION{\bf{The quark-quark scattering amplitude to $O(\alpha_s^2)$
in an arbitary covariant gauge}}
 The process considered is the scattering of two equal mass quarks through
 an angle of 90$^{\circ}$ in their CM system. The lowest order diagram is shown
 in Fig.~1a. The four-vectors of the incoming (outgoing) quarks are $p_1,p_2$
 ($p_3,p_4$). In this configuration the exchanged gluon has a virtuality
 $t = -s/2$ where:
 \begin{eqnarray}
 t & \equiv & (p_1-p_3)^2 = u \equiv (p_1-p_4)^2 ,  \\
 s & \equiv & (p_1+p_2)^2 = (p_3+p_4)^2 .
 \end{eqnarray}
 Denoting the invariant amplitude corresponding to the diagram in Fig.~1a by 
 ${\cal M}^{(0)}$ (quark spin and colour indices are suppressed), then the
 $O(\alpha_s^2)$ amplitude may be written as:
 \begin{equation}
  {\cal M}^{(1)} =  {\cal M}^{(0)} \sum_{i} {\cal A}_i ,
 \end{equation}
 where the one-loop virtual corrections ${\cal A}_i$ are given by the diagrams shown
 in Figs.~1b -- 1i. Three topologically distinct types of diagrams occur: 
 \begin{itemize}
 \item vertex corrections as in Figs.~1b, 1c and the two similar diagrams
 given by the exchange $13 \leftrightarrow 24$;
 \item loop corrections (Figs.~1d--1g);
 \item box diagrams (Figs.~1h, 1i).
 \end{itemize} 
  The leading logarithmic corrections, ${\cal A}_i$, corresponding to 
  the contributions of diagrams that are UV divergent before renormalisation,
   have been derived in a straightforward fashion from the corrections to
  the quark-quark-gluon vertex presented in Ref.~\cite{RDF}. The results, in 
  Feynman gauge, derived using dimensional regularisation~\cite{x6} are 
  presented in Table~1\footnote{In two previous versions of the present
   paper~\cite{Earlyv} the one-loop corrections to the quark-quark scattering amplitude
   were taken
   from Table 1 of Ref.~\cite{x1}. This calculation was performed using 
on-shell
   normalisation~\cite{x5} in which explicit quark and gluon mass parameters
   were introduced. Comparison of the results of Ref.~\cite{x1} with those of
   Refs.~\cite{RDF} and~\cite{x13} and those for the related QED Bhabha 
scattering
   process~\cite{BDH} have revealed a number of inconsistencies with the results
   of Ref.~\cite{x1}. (i) the `Vertex(a + b)' contribution shown is 
   simply the QED result modified by a single multiplicative
   colour factor. The QCD-specific UV divergent term found in 
Refs.~\cite{RDF} and
   \cite{x13} and presented in the first row of Table 1 in the present paper
   is absent. (ii) The `Three Gluon(i+j)' contribution should, according
    to Ref.~\cite{RDF}, be $3y/2$, not $-y$. (iii) Adding up the singly 
logarithmic
   terms of Table 1 of Ref.~\cite{x1} yields, as the coefficient of the logarithm
   of the physical scale, $\beta_0$, but the contribution of `Vertex(a and b)'
    actually corresponds, not to a UV divergent term, but an IR divergent
    one, that is 
    cancelled on adding the contributions of diagrams with real gluon 
    radiation from the external quark lines. Thus Eqs. (2.4) and (2.5) of
    the previous versions of the present paper were incorrect and the 
    appearence of $\beta_0$ in Eq.~(2.12) was fortuitous. However, Eqs. (2.5)
    and (2.6) of the present revised and corrected version are identical
     to Eqs. (2.15) and (2.16) of the previous versions, so that none
     of the subsequent discussion or conclusions are affected by the
     corrections.}.
 Here $Q = \sqrt{-t} = \sqrt{s/2}$ is the physical scale,  and $\mu$ is the 
  renormalistion subtraction scale. $C_A$ is the usual QCD colour factor
  ($C_A$ = number of colours = 3) while $n_f$ is the number of light quark
  flavours contributing to the vacuum polarisation loops in Fig.~1d.
  $\alpha_s^{\mu}$ is the square of the
 renormalised on-shell strong coupling
 constant at the scale $\mu$.
 The corrections ${\cal A}_1-$${\cal A}_4$  may be combined to yield a term
 proportional to the first coefficient in
 the perturbation series in $\alpha_s$ of the beta function of QCD. Denoting
  the quark loop + gluon loop + ghost loop  correction, ${\cal A}_3+{\cal A}_4$, by ${\cal L}^{\mu}$,
  and the vertex correction, ${\cal A}_1+{\cal A}_2$,  by  ${\cal V}^{\mu}$,
  and generalising to an arbitary
 covariant gauge specified by the parameter $\xi$, in which the gluon 
 propagator is written as :
 \begin{equation}
 P^{\mu \nu}(q^2) = - \frac{i}{q^2}\left[ g^{\mu \nu}-(1-\xi)
 \frac{q^{\mu}q^{\nu}}{q^2} \right] 
 \end{equation}
 yields the explicit gauge dependence\footnote{This gauge dependence of one-loop vertex and vacuum
  polarisation insertions containing triple gauge boson couplings is universal for all non-abelian
   gauge theories. See, for example, Ref.~\cite{IZ} Eqs. 12.114 and 12.122.}  of 
${\cal V}^{\mu}$ and ${\cal L}^{\mu}$~\cite{x13}
 \begin{eqnarray} 
 {\cal V}^{\mu}(\xi) & = &  -\frac{\alpha_s^{\mu}}{\pi} \frac{3}{4} (3 +\xi)
  \ln (\frac{Q}{\mu}) ,  \\
{\cal L}^{\mu}(\xi) & = &  \frac{\alpha_s^{\mu}}{\pi}\left[-\frac{3}{4}
(\frac{13}{3}-\xi) + \frac{n_f}{3}\right]\ln(\frac{Q}{\mu}) .
\end{eqnarray}
   
\par  Adding Eqs. (2.5) and (2.6):
\begin{equation}
{\cal V}^{\mu}(\xi)+{\cal L}^{\mu}(\xi) = -\frac{\alpha_s^{\mu}}{\pi}(\frac{11}{2}
  - \frac{n_f}{3}) \ln(\frac{Q}{\mu}) =
-\frac{\alpha_s^{\mu}}{\pi} \beta_0 \ln(\frac{Q}{\mu}) ,
\end{equation}
where $\beta_0$ is the first coefficient of the QCD $\beta$ function 
 [ $a_s \equiv \alpha_s/\pi$] :
\begin{equation} 
\mu \frac{\partial a_s}{\partial \mu} = \beta(a_s) = -\beta_0 a_s^2 + ...
\end{equation}
 To $O(\alpha_s^2)$, $\alpha_s^{Q}$ may be identified with the solution,
 $\alpha_s^{RGE}(Q)$, 
 of the RGE (2.8):
 \begin{eqnarray} 
 {\cal M}^{LO} & = & {\cal M}^{(0)} + {\cal M}^{(1)} = {\cal M}^{(0)}
 (1+{\cal V}^{\mu}(\xi)+ {\cal L}^{\mu}(\xi)) \nonumber \\
 & = & \frac{{\cal M}^{(0)}}{\alpha_s^{\mu}}\left[\alpha_s^{\mu}
 [1- \frac{\alpha_s^{\mu}}{\pi}\beta_0 \ln(\frac{Q}{\mu})]\right]
 \nonumber \\
 & = & \frac{{\cal M}^{(0)} \alpha_s^{RGE}(Q) }{\alpha_s^{\mu}}+
 O\left( (\alpha_s^{\mu})^2\right) ,
\end{eqnarray} 
where
\begin{equation}
 \alpha_s^{RGE}(Q) = \alpha_s^Q = \frac{ \alpha_s^{\mu}}
{1+ \frac{\alpha_s^{\mu}}{\pi}\beta_0 \ln(\frac{Q}{\mu})} ,
\end{equation}
${\cal M}^{LO}$ denotes the amplitude including Leading Order
vertex and vacuum polarisation corrections. In the following section the
amplitude ${\cal M}^{(\infty)}$, in which these corrections are summed to
all orders in $\alpha_s$, is derived.
\par It can be seen that, at one loop, the gauge invariant result (2.7) is
  obtained on adding the UV divergent contributions of diagrams (b)--(g) in Fig.~1. 
  The box diagrams (h) and (i) and those obtained by exchanging the internal
  quark and gluon propagators, form, together with the diagrams where a 
  single gluon is radiated from one of the external quark lines of Fig.~1a,
  another gauge invariant set. Indeed the contribution to the cross section
  of this set of diagrams, which involve only abelian quark-gluon 
  couplings, is obtained by multiplying
  the result for the analogous QED t-channel Bhabha scattering
  process~\cite{BDH} by the appropriate QCD colour factor:
  $C_2(N_c) = (N_c^2-1)/2N_c = 4/3$. The contribution of the
   box diagrams alone is UV finite, but IR divergent for the case 
   of massless gluons. This IR divergence is cancelled
   on adding the contribution of the real gluon radiation
  diagrams. The contributions of the box diagrams and the
  real radiation diagrams are however, separately, 
gauge invariant\footnote{
 See Ref.~\cite{x14} for a complete diagrammatic discussion of gauge 
cancellations
 in the one-loop corrected quark-quark scattering amplitude.}.
 Since the box diagrams are UV finite, and so do not require renormalisation,
 they do not contribute to the QCD effective charge. All the terms that
 contribute to the latter at one loop are presented in Table 1; 
they are UV divergent before renormalisation.

\SECTION{\bf{The resummed quark-quark scattering amplitude}}
The topographical structures\footnote{The use of the 
word `topographical structure' to indicate
    a particular disposition of vertex and (possibly resummed) self-energy
    insertions in a diagram is deliberate. Diagrams with internal lines in
    the vertex or self-energy insertions have a different topology but the same
    topography as the one-loop diagrams shown in Fig.~2.} of the diagrams that modify the gluon propagator
in the quark-quark scattering amplitude at $O(\alpha_s^2)$, $O(\alpha_s^3)$
and $O(\alpha_s^4)$ are shown in Figs. 2a,b,c respectively. $V$ and $L$ denote
vertex and loop (vacuum polarisation) contributions:
\begin{eqnarray}
 V & = & V_1+ V_2 , \\
 L & = & \sum_{i=1}^{n_f} F_i + G_1 +G_2 +G_3 .
\end{eqnarray}
$V_1$, $V_2$  correspond to the diagrams in Fig.~1b, 1c ;  
$\sum_{i=1}^{n_f} F_i$
to Fig.~1d and $G_1$, $G_2$, $G_3$ to Figs.~1e,f,g.
  In the case that only one vertex insertion occurs there is a factor 2 for
the two ends of the gluon propagator. Since the propagator has only two ends
the vertex corrections are never higher than quadratic in the perturbation
series for the amplitude. Although the topographical structure of diagrams
containing vertex corrections is different at $O(\alpha_s^3)$ than at
$O(\alpha_s^2)$ it remains the same at all higher orders. The all--orders
resummed amplitude is:
\begin{eqnarray}
{\cal M}^{(\infty)} & = & {\cal M}^{(0)}\left[ 1 + 2V + L \right. \nonumber \\
                  &   &~~~~~+V^2 +2VL + L^2  \nonumber \\ 
                  &   &~~~~~\left.+V^2L +2VL^2 + L^3 + ...~~ \right]  \nonumber \\ 
                  & = & {\cal M}^{(0)}\left[ 1 + L + L^2 + ... +V^2 \right.
                        (1 + L + L^2 + ... \nonumber \\  
                  &   &~~~~~\left.+2V(1 + L + L^2 + ...~~ \right]  \nonumber \\ 
                  & = & \frac{{\cal M}^{(0)}(1+V)^2}{1-L} .
\end{eqnarray}  
In an arbitary covariant gauge with momentum subtraction at scale $\mu$,
and in leading logarithmic approximation, 
$V = {\cal V}^{\mu}(\xi)/2$, $L = {\cal L}^{\mu}(\xi)$ so that                   
\begin{equation}
{\cal M}^{(\infty)} = \frac{{\cal M}^{(0)}  \alpha_s^{eff}(Q)}{\alpha_s^{\mu}}
= \frac{{\cal M}^{(0)}(1+\frac{1}{2}{\cal V}^{\mu}(\xi))^2}{1-{\cal L}^{\mu}(\xi)}
\end{equation}
leading to the following expression, in leading logarithmic approximation, for the
 resummed one-loop 
effective charge:
\begin{equation}
\alpha_s^{eff}(Q) = \alpha_s^{\mu}\frac{\left[1-\frac{3 \alpha_s^{\mu}}{8 \pi}
(3+\xi) \ln(\frac{Q}{\mu})\right]^2}
{1+\frac{\alpha_s^{\mu}}{4 \pi}\left[13-3\xi-\frac{4 n_f}{3}\right]
\ln(\frac{Q}{\mu})} .
\end{equation}
The conventional one-loop QCD running coupling constant Eq.~(2.10)
 is recovered only for the special choice of gauge parameter $\xi = \xi_L=-3$
 (`loop gauge'). Only in this case does $\alpha_s^{eff}(Q)$ decrease monotonically
  with increasing $Q$. For any other choice of gauge $\alpha_s^{eff}(Q)$ does not
  show `asymptotic freedom' as $Q \rightarrow \infty$ but instead {\it increases}
  as $\ln Q$ when $\xi \ne \xi_V$ and as $\ln^2 Q$ when $\xi = \xi_V$. The
   choice $\xi = \xi_V$ (`vertex gauge') where
   \[ \xi_V = (39-4 n_f)/9  \]
   corresponds to a vanishing coefficient of $\ln(Q/\mu)$ in the denominator
 of Eq.~(3.5). As discussed in more detail below, only in loop gauge is the
 equation for the effective charge `self similar' like the effective charge in
 QED or the solution (2.10) of the RGE Eq.~(2.8). Even in loop gauge the
  effective charge of Eq.~(3.5) does not decrease without limit as 
$Q \rightarrow \infty$. The maximum possible scale $Q_L$ (Landau scale) is
determined by the convergence properties of the geometric sum that yields the
denominator of Eq.~(3.5). The geometric series is convergent  
provided that $|L|<1$~\cite{x15}. This implies that:  
\begin{equation}
\left|\frac{\alpha_s^{\mu}}{4 \pi}[13-3 \xi- \frac{4 n_f}{3}]
\ln(\frac{Q}{\mu})\right| < 1 .
\end{equation}
The corresponding Landau scale is then:
\begin{equation}
Q_L = \mu \exp \left[\frac{4 \pi}{\alpha_s^{\mu}[13-3 \xi- \frac{4 n_f}{3}]}\right]
\end{equation} 
and
\begin{equation}
\alpha_s^{eff}(Q_L) = \frac{\alpha_s^{\mu}}{2}
\left[1-\frac{3 \alpha_s^{\mu}}{8 \pi}
(3+\xi) \ln(\frac{Q_L}{\mu})\right]^2 .
\end{equation}
For $\xi = -3$:
\begin{equation}
\alpha_s^{eff}(Q_L) = \frac{\alpha_s^{\mu}}{2} ,
\end{equation}
so the convergence condition (3.6) implies that in this case $\alpha_s^{eff}$
cannot evolve down by more than a factor of $\frac{1}{2}$ of its initial value
before Eq.~(3.5) diverges
--- there is no `asymptotic freedom'. The convergence conditions for a geometric series
 with a negative common ratio, such as that which generates the QCD running coupling constant,
 are derived below in an Appendix.
 Unlike in the case of the Landau pole~\cite{Landau} of QED, with a positive common ratio, $r = |r|$,
 where the running coupling constant is proportional to $1/(1-|r|)$, it is not 
   obvious, by inspection, that the QCD perturbation series is {\it not} equal to $1/(1+|r|)$ when  $|r|>1$. However
   (see the Appendix) this is quite clear from the exact formula, Eq.~(A1), giving the sum for any finite number of terms. Physicists have the right to be as free as possible
    in making conjectures in their attempts to describe nature in the simplest way possible, but not
    to make ones that are contrary to mathematical laws. 
\par Numerical values of $Q_L$ for $\mu = 5$ GeV,  $\alpha_s^{\mu} = 0.2$
 and $n_f=5$ (corresponding, approximately, to the experimental
 value of $\alpha_s^{\mu}$)  
 and four different choices of gauge parameter are presented in Table 2.
 For loop gauge ($\xi = -3$) $Q_L$ is only 300 GeV, with phenomenological 
 consequences perhaps already at the Fermilab $ \rm{p} \overline{\rm{p}}$ collider, 
 but certainly at the future LHC pp collider. It may be noted that, in all
 cases except vertex gauge, $Q_L$ lies well below the Grand Unification (GUT)
 scale of $\simeq 10^{15}$ GeV. Since for any gauge choice except $\xi = -3$,
 $\alpha_s^{eff}$ diverges as $\ln Q$ or $\ln^2 Q$ at large $Q$, there can be
 no `unification'~\cite{x16,x17} of the strong and electromagnetic interactions
 at large scales $Q$, for any choice of gauge parameter, at least if the running
 strong coupling constant is identified with an effective charge such as that
 in Eq.~(3.5). In fact all studies, to date, of Grand Unification have implicitly
 used loop gauge where the maximum scale $Q_L$ is only $\simeq$ 300 GeV.
  In vertex gauge, since all vacuum polarisation
 contributions vanish, there is no convergence limitation on $Q$ in Eq.~(3.5).

   \par The behaviour of  $\alpha_s^{eff}$ at large values of $Q$ depends
    on the value of the Landau scale and value, $Q_0$, of $Q$ at which the numerator
    of Eq.~(3.5) vanishes:
 \begin{equation}
Q_0 = \mu \exp \left[\frac{8 \pi}{3 \alpha_s^{\mu}(3+\xi)}\right] .
\end{equation} 
  The value of $\xi$, $\xi_{0}$, at which $Q_L$ and $Q _0$ are equal is given by (3.7) and (3.10)
  as:
 \begin{equation}
 \xi_{0}(n_f)  = \frac{1}{9}\left(17-\frac{8 n_f}{3}\right)
\end{equation} 
  For $n_f = 5$, $\xi_{0}= 0.4074...$. When $\xi < \xi_{0}$, (e.g. Loop and Landau gauges) and
  $Q_L <  Q_0$, then  $\alpha_s^{eff}$ diverges before vanishing and so decreases monotonically with increasing $Q$
    within its convergence radius. For $\xi > \xi_{0}$ (e.g. Feynman and Vertex gauges)  $Q_L >  Q_0$.
    In this case  $\alpha_s^{eff}$ first decreases with increasing $Q$, reaching a minimum value at zero, and
     then increases with increasing $Q$ up to $Q_L$. In vertex gauge  $Q_L$ is infinite.
  Numerical values of $Q_0$ for the
  same parameter choices as in Table 2 are presented in Table 3. The scale $Q_0$
  is defined (i.e.  $Q_L >  Q_0$.) only for Feynman gauge ($Q_0 = 176$ TeV) and
   Vertex gauge ($Q_0 = 18.1$ TeV).  
\par The diagrams shown in Fig.~2 each contain only a single `dressed' virtual
  gluon. It is also possible to consider cases in which the subsitution of the
  series of diagrams illustrated in Fig.~2, and summed in Eq.~(3.3), is made
  in a diagram containing more than one virtual gluon (for example the
  box diagrams shown in Fig.~1h and 1i). This will result in a further
  higher order correction to the quark-quark scattering process. It is clear,
  however, that the
   manifest gauge
  dependence of the resummed one-loop contribution will be unaffected by 
  the presence of other virtual gluons
  (`dressed' or not) in the diagram. Every virtual gluon has only
   two ends, and so the resummed vertex correction is at
   most quadratic. Then no cancellation is possible of the
  gauge dependent pieces of multiple vacuum polarisation
  insertions. However, as further discussed in Section~5
  below, the box diagrams themselves can be resummed. This typically results
  in a Sudakov-like double logarithm, not a geometric series as found
  for resummed vacuum polarisation diagrams.

\SECTION{\bf{Self-Similarity Properties of $\alpha_s^{eff}$ and the
Renormalisation Group}}
Introducing the abbreviated notation:
\[ v(\xi) \equiv \frac{3}{8} (3 + \xi)~~~,~~ l(\xi) = \frac{1}{4}\left[
13-3\xi-\frac{4 n_f}{3}\right] \] 
\[ \alpha_s^{eff}(Q)/\pi \equiv a_Q~~~,~~ \alpha_s^{\mu}/\pi \equiv a_{\mu}~~~,
~~\lambda \equiv \ln (Q/\mu) \]
Eq.~(3.5) may be written as:
\begin{equation}
a_Q = \frac{a_{\mu}[1-a_{\mu}v(\xi)\lambda]^2}{1+a_{\mu}l(\xi)\lambda} .
\end{equation}
With $\xi = -3$, $v(-3) = 0$, Eq.~(4.1) becomes 
\begin{equation}
a_Q = \frac{a_{\mu}}{1+a_{\mu}l(-3)\lambda}.
\end{equation}  
Eq.~(4.2) is the solution of a one-loop RGE similar to that for the QED 
effective charge~\cite{x10,x11}:
\begin{equation}
\frac{Q}{a_Q}\frac{\partial a_Q}{\partial Q} = -l(-3) a_Q = -\beta_0 a_Q ,
\end{equation}
where
\begin{equation}
\beta_0 = l(-3) = \frac{11}{2}-\frac{n_f}{3}.
\end{equation}
 For a gauge choice such that $v \ne 0$ the partial differential equation
  satisfied by $a_Q$ is:
\begin{equation}
\frac{Q}{a_Q}\frac{\partial a_Q}{\partial Q} = -a_Q \left[\frac{l(\xi)}
{(1-a_{\mu}v(\xi)\lambda)^2}+
2\frac{v(\xi)(1+a_{\mu}l(\xi)\lambda)}{(1-a_{\mu}v(\xi)\lambda)^3}\right] ,
\end{equation}
so that, in this case, the effective charge $a_Q$ does not satisfy the RGE 
 (4.3).
Expanding in powers of $a_{\mu}$ on the right hand side of Eq.~(4.5) gives:
\begin{eqnarray}
\frac{Q}{a_Q}\frac{\partial a_Q}{\partial Q}& = &-a_Q \left[l(\xi)+2 v(\xi)
+ O(a_{\mu})\right] \\
& = &-a_Q \left[\beta_0
+ O(a_{\mu})\right] .
\end{eqnarray}
Thus, neglecting terms of $O(a_Q a_{\mu}) \simeq O(a_{\mu}^2)$ so that only
 the first term in the QCD perturbation series is retained, $a_Q \rightarrow
 a_Q^{(1)}$ and it can be seen that $a_Q^{(1)}$, for an arbitary gauge choice,
 satisfies the same partial differential equation as the one-loop 
 all orders resummed $a_Q$ in loop gauge. The relation of this result to 
 previous derivations of the QCD running coupling constant, where it has 
 generally been conjectured that a gauge invariant result is obtained to all
 orders in perturbation theory, is discussed in the following Section.
 \par The equation (4.2) is self-similar in the sense that , for any values of
 $\mu$ and $Q$ the equation defined by the exchange $\mu \leftrightarrow Q$
 is identical to the original equation. A consequence of this symmetry 
 property is that $a_Q$ in Eq.~(4.2) is independent of $\mu$ (with the 
 important caveat that, since the denominator is the sum to infinity of a
 geometric series, $\mu$ must be such that $|a_{\mu}l(-3)\lambda|<1$),
 and that in the equation with $\mu \leftrightarrow Q$, $a_{\mu}$ is 
 independent of Q. This is the mathematical basis of the Renormalisation
  Group~\cite{x7,x8,x9}.
 Such a universal self-similarity property is not, however, shared by Eq.~(4.1)
 when $v(\xi) \ne 0$. 
 \par Eq.~(4.1) is self-similar under the exchange $\mu \leftrightarrow Q$
 provided that the equation: 
\begin{equation}
a_{\mu} = \frac{a_{Q}[1+a_{Q}v(\xi)\lambda]^2}{1-a_{Q}l(\xi)\lambda}
\end{equation}
and Eq.~(4.1) are both valid.
Simultaneous solution of Eqs. (4.1), (4.8) in the case that 
$v(\xi) \ne 0$ leads
to a quadratic equation for $a_{\mu}$ with the solution:
\begin{equation}
a_{\mu} = \frac {1}{v(\xi)\lambda}+
\frac{a_Q}{2}\pm\frac{1}{2}\sqrt{(\frac{2}{v(\xi)\lambda}+a_Q)
(\frac{2}{v(\xi)\lambda}-3a_Q)}.
\end{equation}
Real solutions of Eq.~(4.9) exist provided that either
\begin{equation}
\frac{2}{3v(\xi)\lambda}> a_Q > -\frac{2}{v(\xi)\lambda}
\end{equation}
(both factors under the square root positive) or
\begin{equation}
\frac{3v(\xi)\lambda}{2}> \frac{1}{a_Q} > -\frac{v(\xi)\lambda}{2}
\end{equation} 
(both factors under the square root negative). 
\par The solution for $a_{\mu}$, Eq.~(4.9), is independent of $l(\xi)$. Thus for fixed
values of $a_Q$, $Q$, $\mu$, $v(\xi) \ne 0$, satisfying the conditions (4.10) or
(4.11) there are, in general, two values of $a_{\mu}$ such that Eq.~(4.1) is
self-similar. This value of $a_{\mu}$ has however no relation to the effective
charge at the scale $\mu$ given by Eq.~(4.1) when $Q=\mu$. Choosing $Q = 90$ GeV
and Landau gauge then (see Fig.~3) $a_Q = 0.095/\pi = 0.0302$. The choice 
$\mu = 5$ GeV gives $2/(v(0)\lambda) = 1.933$. The $a_Q$ terms under the square root
 of Eq.~(4.9) may then be neglected, leading to the solutions:
 \[ a^+_{\mu} \simeq \frac{2}{v(0)\lambda}+\frac{a_Q}{2} = 1.9648 \]
 \[ a^-_{\mu} \simeq \frac{a_Q}{2} = 0.0152 \]
  to be compared with the physical value:
  \[ a_{\mu} = \alpha^{eff}(5{\rm GeV})/\pi =0.2/\pi = 0.064 \]
 \par Unlike for the special case $v = 0$, the first derivative of $a_Q$ in 
 general depends on the scale $\mu$. For $Q=\mu$ the derivative 
 is a negative constant 
 fixed by the first coefficient in the perturbation series for the beta function
 as in Eq.~(4.3). For any other choice of $\mu$ when $v \ne 0$ the derivative
 varies with $Q$ and $\mu$ according to Eq.~(4.5). Thus, if the effective charge is 
 parametrised in terms of $a_{\mu}$, $a_{\mu'}$ at the reference scale $Q_R$:
\begin{equation}
a_{Q_R} = a_{\mu}\frac{[1-a_{\mu}v(\xi)\lambda_R]}{[1+a_{\mu}l(\xi)\lambda_R]}
= a_{\mu'}\frac{[1-a_{\mu'}v(\xi)\lambda_R']}{[1+a_{\mu'}l(\xi)\lambda_R']},
\end{equation}
\[ \lambda_R \equiv \ln(Q_R/\mu)~,~ ~\lambda_R' \equiv \ln(Q_R/\mu') \]
then for $Q \ne Q_R$ and $v(\xi) \ne 0$ the effective charge $a_Q$
 predicted by the formula containing $\mu$ (the second member of Eq.~(4.12))
 will be different to that predicted by that containing $\mu'$ (the third member
 of Eq.~(4.12)), so that the value of $a_Q$ depends upon the choice of renormalisation scale
  --- it is no longer invariant. 
 Renormalisation group invariance with respect
  to the choice of the scale $\mu$ is
 therefore not respected unless $v(\xi)=0$, i.e. $\xi =-3$.
    
\SECTION{\bf{Discussion}}
 In the original derivations of the `asymptotic freedom' property of QCD
 ~\cite{x18,x19} no calculations were performed beyond the lowest non-trivial
 order, $O(\alpha_s^2)$, and no actual amplitudes for physical processes were 
 considered. It was conjectured (without any check by direct diagrammatic
 calculation) that the QCD running coupling constant (RCC) could, in general
 (for any choice of gauge) be identified with the solution of the differential
 equation (2.8). The one-loop  QCD beta function was calculated by considering
 the Callan-Symanzik~\cite{x10,x11} equation for an irreducible $n$-point function
 (typically the gluon-quark-quark vertex or the triple gluon vertex). Calculation
 of the anomalous dimensions of the quark and gluon fields then yields the
 (gauge invariant) expression for the first beta function coefficient $\beta_0$
 in Eq.~(2.8) above. An analogous result is obtained above by considering
 the unresummed  one-loop correction to the physical quark-quark scattering
 amplitude. The true high order behaviour, however, corresponds to the sum
 of {\it all} the possible  amplitudes for the process of interest.
 For this the actual topographical structure of the diagrams contributing
 to the amplitude must be properly taken into account. 
  Renormalisation scale invariance of the RCC, and the asymptotic freedom
   property, for an arbitary choice of covariant gauge,
   are not confirmed by the 
 diagrammatic calculation of higher order corrections in the case of the
  quark-quark scattering amplitude considered in this paper. 
  \par Gauge dependence of the RCC in QCD has been considered previously in the literature,
  but usually as an effect only at the two-loop level and at higher orders. It was pointed
  out that, in the case when the bare parameters of the theory are held fixed, the 
  gauge parameter becomes scale dependent, and for certain momentum subtraction
  renormalisation schemes, the second coefficient of the beta function is both
  scheme and gauge dependent~\cite{x20}. For an arbitary covariant gauge specified
  by the fixed parameter $\xi$, as in the one-loop discussion above, the second
  beta function coefficient is, however, both gauge and renormalisation scheme
  invariant. Assuming that the RCC, mass and gauge parameter each satisfy
  renormalisation group equations similar to (2.8), and solving the coupled
  system of differential equations, solutions were found for the RCC that
  strongly depended on the initial conditions imposed on the running gauge
  parameter~\cite{x20}. These solutions exhibit either asymptotic freedom-like behaviour
  or increase with increasing scale until an ultraviolet fixed point is 
  reached~\cite{x21,x22}. As commonly done in the literature, the RCC was treated as an
  independent mathematical object, without reference to any actual physical process,
  and the renormalisation group equations were assumed to hold without specific
  diagrammatic justification.
  It is shown above that, if the RCC is identified with the effective charge
  of the quark-quark scattering amplitude, 
the one-loop renormalisation group equation\footnote{In 
Refs.~\cite{x21,x22} the one-loop RGE was assumed to be gauge independent
  and given by the conventional formula (2.10).}
  holds only for the specific gauge choice $\xi = -3$. The gauge parameter can 
  then neither vary nor satisfy a RGE.
 \par When quark mass effects are taken into account, the one-loop beta function 
  coefficient is also gauge dependent, and has a value which depends on the
  particular $n$-point Green's function considered for its derivation. The mass
  dependent corrections to the triple gluon vertex~\cite{x23} and the 
  gluon-ghost-ghost vertex~\cite{x24} are different. A detailed discussion may
  be found in Ref.~\cite{x25}. A corollary is that the RCC in physical amplitudes
  is both gauge and process dependent at physical scales where quark mass effects
  (other than those contained in the asymptotically dominant logarithmic
  terms) are important. 
  \par The effective charge (3.7) has been calculated here for the simple case
  of quark-quark scattering with a unique physical scale $Q =\sqrt{-t}$.
   In this case the direct physical interpretation as the strength
    of the interaction between two
  currents varying as a function of their separation ($\simeq 1/Q$) is 
  particularly transparent.
   However, since every dressed propagator has just two ends, similar 
   expressions for the RCC (in general a function of some running
   loop 4-momentum $k$) are expected in all physical amplitudes
   containing virtual gluon lines.
   Two examples are shown in Fig.~4. Fig.~4a shows the three topographically
   distinct classes of diagrams that contribute to the anomalous magnetic
   moment of a heavy quark at $O(\alpha_s^3)$. In Fig.~4b the same classes
   of diagrams are shown for the four-loop photon proper self energy function
   due to radiatively corrected quark vacuum polarisation loops. As for the
   quark-quark scattering case the same topographical structures (giving at most
   a quadratic dependence on the vertex corrections) is found at all higher orders
   in the `dressed gluon propagator'. The diagrams of Fig.~4b are related via the
   optical theorem and analytical continuation to the process: 
   \[ e^+e^- \rightarrow \gamma^* \rightarrow q \overline{q}+X \]
   where $X$ denotes $g$, $gg$ or $q' \overline{q}'$.
   \par There has been considerable recent interest in the structure, in high orders
   of perturbation theory, of diagrams containing chains of vacuum polarisation loops
   in internal gluon propagators (for example the generalisation to higher orders
   of the $O(\alpha_s^3)$  diagrams with two vacuum polarisation loops shown in 
   Fig.~4). The anstatz used for these so-called `renormalon chains'
   ~\cite{x26} is to replace $n_f/3$ in a calculation considering only $n_f$ different
   flavours of fermion vacuum polarisation loops
    with $n_f/3-11/2= -\beta_0$. This is
   clearly a good approximation in the limit $n_f \rightarrow \infty$. As inspection
   of Fig.~4 shows, however, this will result, in any gauge in which the vertex correction
   is non-vanishing, in a miscounting of the contribution 
   of the latter, which are included at order
   $n$ in terms of the form $\beta_0^n$, but actually should never appear at higher order than
   quadratic in the perturbation series. With however the gauge choice 
   $\xi = -3$ (loop gauge) all vertex corrections vanish and the renormalon
   chains are correctly given by the above replacement. This gauge choice is,
   in any case, the one universally (although tacitly) made in all phenomenological
   applications of the QCD running coupling constant, where renormalisation
   group invariance is assumed.
   Thus the `Naive Non-Abelianization' (NNA) ansatz~\cite{x27} or `Large $\beta_0$ Limit'
   that is typically assumed in phenomenological studies of renormalon effects is 
   a correct one only in loop gauge.   
  The remark that the choice of gauge $\xi = -3$
   implies the vanishing of all vertex corrections, has previously been made in the 
   context of two-loop corrections to heavy quark production in 
   $e^+e^-$ annihilation~\cite{x28}.
   \par An instructive example of the inconsistent treatment of Feynman diagrams by the
      NNA anstatz as well as a illustration of the gauge dependence of fixed order perturbative
    QCD calculations beyond next-to-leading-order (NLO) is provided by the analysis of the moments
    of non-singlet anomalous dimensions of deep-inelastic nucleon structure functions in
     Ref.~\cite{Mikhailov}. In this work the effect of insertion of an arbitary number of
     quark or gluon vacuum polarisation insertions in the virtual gluon propagator of
      a forward Compton scattering amplitude (or the equivalent diagrams in the Operator Product
      Expansion, as in Fig.~1 of Ref.~\cite{Mikhailov}) is considered. An example of an O($\alpha_s$)
      diagram that contributes to this amplitude is shown in Fig.~5a. One and two vacuum polarisation
       loop insertions as considered in  Ref.~\cite{Mikhailov} are shown in the fourth topographical
    diagrams of Fig.~5b and 5d respectively. The predictions of this procedure
     for the loop ($\xi = -3$) and Landau ($\xi = 0$) gauges were compared to the exact massless
     next-to-next-to-leading-order (NNLO) calculation in Feynman gauge\footnote{The formulae
      given in Ref.~\cite{LRV} are actually in Landau gauge. However, it is stated in the paper
      that for low order moments $n =2,4$ ``...the diagrams were run with a gauge parameter $\xi$
       in the gluon progagator $g_{\mu\nu}-\xi q_{\mu}q_{\nu}/q^2$.'' The assumed value
       (or values) of their parameter $\xi$ ($1-\xi$ in the notation of the present paper) were
      not stated and for the moments $n =6,8$ the gauge parameter was not included (presumably
     it was set to zero in the calculations) which corresponds to choosing Feynman gauge. It is assumed
     here that results compatible with this choice of gauge were also obtained for the
     $n =2,4$ moments.} ($\xi = 1$) of Ref.~\cite{LRV}.
     Table 4, extracted from
     Table 1 of  Ref.~\cite{Mikhailov}, shows the results of this comparison for the $n = 2$ moment 
     at NLO, as in Fig.~5b, and NNLO, as in Fig.~5c and 5d. In the calculations
     of Ref.~\cite{LRV} the contributions of all 353 Feynman diagrams contributing to the anomalous
      dimensions up to NNLO were evaluated, whereas in Ref.~\cite{Mikhailov} only quark and gluon
      loop vacuum polarisation insertions in Fig. 5a and the other Compton scattering diagrams 
       with one virtual gluon line (see for example Fig.~1 of Ref.~\cite{MNS}) were considered.
     Even so, agreement is found with the Loop gauge renormalon calculation
     at the 20$\%$ level at NLO for the $n = 2$ moment. Even better
      agreement is found for higher moments ---it is good at 4.7 $\%$ for the  $n = 6$ moment.
     However, at NNLO no agreement is found; indeed predictions of the `renormalon dominated'
    approximation with either choice of gauge
     have even a different sign to the exact NNLO calculation. The reasonable agreement at NLO
    between the two calculations when loop gauge is employed in Ref.~\cite{Mikhailov}
    can be understood by inspection of Fig.~5b. This choice
    of gauge is equivalent, at NLO, to performing a calculation in an arbitary covariant gauge,
   in which the non-abelian one-loop corrections to the quark-gluon coupling are included as well as
    vacuum polarisation insertions.
   All the NLO vertex and vacuum polarisation corrections are included in the topographical
   diagrams of Fig.~5b. The sum of these contributions is gauge invariant. The situation 
   is quite different at NNLO. The subset of NNLO diagrams shown in Fig.~5c, obtained from
    those of Fig.~5b by inserting an additional virtual gluon between the incoming and outgoing
    quark lines also gives a gauge invariant result and is correctly described in loop gauge.
     This is not the case for the diagrams of Fig.~5d ---the corresponding sum of the diagrams 
     is of the form $V^2 +2VL +L^2$, which is manifestly gauge dependent. Other NNLO contributions
     with the same colour factors as in Table 4 arise from irreducible two-loop vertex and
     vacuum polarisation diagrams.  The topographical pattern is the same as in 
      Fig.~5b with the one-loop insertions replaced by irreducible two-loop insertions.
      As discussed in more detail below, if these two-loop insertions satisfy a Ward identity similar
      to that respected by
      the one-loop insertions, this contribution will also be gauge invariant. No possibility
      exists however to remove the gauge dependence of the contribution of the diagrams 
      shown in Fig.~5d. This explains the breakdown of gauge invariance shown by the
       $C_FC_A^2$ NNLO entries of Table 4.
       \par In an attempt to give a diagrammatic justification of the NNA ansatz, Beneke in 
      Ref.~\cite{x26} considered NLO loop corrections as in Fig.~2a or Fig.~5b above, for the case of
      quark pair production by a vector current. Performing the calculation in Landau gauge it was found
      that the beta function of the corresponding QED calculation was replaced by the one-loop
   QCD beta function. If this calculation had been performed in an arbitary covariant gauge,
      the gauge invariance of the QCD beta function would have been demonstrated. This in no way
      justifies the general use of the NNA ansatz since manifest gauge dependence as in Fig.~2b or Fig.~5d
      first occurs at NNLO where the gauge dependence of the $V^2$ term is not cancelled by
      NNLO vacuum polarisation contributions.
      \par In connection with the work presented in the present paper it is important to notice 
     that the renormalon singularities arise due to loop integrals over the virtuality of an
      internal photon or gluon line in a Feynman diagram, as in Figs.~4 and 5. Although 
     the renormalon singularities are related to singular IR or UV  behaviour of the running
     coupling constant ---as discussed by Lautrup in Ref.~\cite{x26}, in QED it is the UV
     Landau singularity--- the renormalon singularity occurs for arbitary values of the
     external physical scale due to the infinite range of the internal loop 
     momentum\footnote{In a discussion of renormalons in a review talk by S.~Forte~\cite{Forte} the following
      important statement can be found: `If the series had alternating signs the singularity
      (ultraviolet renormalon) would not be on the path of the integral' (in the Borel transform)
        'but the integral would still run outside the radius of convergence of the series;
        we will not discuss this any further.' This is the only place in the literature, to the 
       present writer's knowledge, where the limited domain of convergence, in the UV limit, of the
         QCD RCC (discussed in detail in the present paper) is mentioned. The perturbation series
     corresponding to the RCC of QCD indeed has `alternating signs'.}.
     In contrast, in the simple case of quark-quark scattering discussed in the present paper,
    there is no integration over the virtuality of the gluon line in which
     the vertex and loop corrections are inserted and the scale in the running coupling constant is
    identical to the physical scale of the problem. The diagramatic analysis 
     shown in Fig.~2 is therefore much simpler and the breakdown of gauge invariance appears as
     soon as the NNLO contributions of Fig.~2b are evaluated.

   \par The classification of diagrams as in Figs.~2, 4 and 5, according to the 
   categories (in an obvious notation) $L^n,~VL^n,~V^2L^n$, will remain when
   $L$, $V$ are calculated with an arbitary number of internal lines. The global
   structure of Eq.~(3.5) will then remain the same for calculations including an
   arbitary number of loops, though additional non-leading terms in $\ln Q$ will
   result from integration over internal loops in the basic one-loop 
   vacuum polarisation and vertex diagrams shown in Fig.~1.
   \par As in Ref.~[1], only the leading logarithmic terms in the one-loop
   correction (and hence in the resummed effective charge) have been taken into
   account in the above discussion. Constant terms in $V$ and $L$ 
   have been neglected. For a general renormalisation scheme however, constant
   gauge and renormalisation scheme dependent terms also occur, so that
   Eq.~(4.1) is replaced by the expression:
   \begin{equation}
   a_Q = a_{\mu}\frac{\{1-a_{\mu}[v(\xi)\lambda+c_v(\xi)]\}^2}
   {1+a_{\mu}[l(\xi)\lambda+c_l(\xi)]} .
   \end{equation}
   In the $\overline{\rm MS}$ scheme~\cite{x14}:
   \[ c_l(\xi) =
    \frac{10 n_f}{9}-\frac{97}{12}-\frac{3}{8}\xi -\frac{3}{16} \xi^2 \]
    By a suitable scale choice $\mu = \mu'$ and with $\xi = -3$ Eq.~(5.1) may
    be written as:
   \begin{equation}
   a_Q = \frac{a_{\mu'}}
   {1+a_{\mu'}[l(-3)\ln\frac{Q}{\mu'}+c_l(-3)]} .
   \end{equation}
   So in this (non asymptotic) case even in loop gauge the effective charge
   does not correspond exactly to  the solution (2.10) of the one-loop 
   RGE (2.8). Numerically $l(-3) = 3.833$, $c_l(-3) = -3.090$ for 
   $n_f = 5$. Thus, in the $\overline{MS}$ scheme in loop gauge, the
   resummed one-loop
   invariant charge is only asymptotically a renormalisation group invariant
   when constant terms in the one-loop correction are retained.
   \par Following the observation that the UV divergent parts of the vertex 
   corrections in Figs.~1b and 1c may be associated with related diagrams in which
   the virtual quark propagators are shrunk to a point (or `pinched') it was 
   suggested~\cite{x29,x30} to redefine a gluon proper self energy function
   by adding to the contributions of Figs.~1d--1g, that of the pinched
   vertex diagrams. At one-loop order the resulting gluon proper self energy
   function is then gauge invariant. It was then (incorrectly) stated that
   a gauge invariant resummed gluon propagator may be trivially derived
   from the one-loop result (for example, Eq.~(2.19) of Ref.~\cite{x30}). 
   In fact it is easy to show, quite generally, that if the one-loop corrected quark-quark 
   scattering amplitude is gauge invariant (the correct initial assumption
   of the pinch technique calculations of Refs.~\cite{x29,x30}) then resummed 
   amplitudes at all higher orders must be gauge dependent.
   Introducing the gauge invariant one-loop quantity:
   \begin{equation}
   B \equiv L + 2V .
   \end{equation}
   The resummed amplitude at O($\alpha_s^{n+2}$) may then be written for 
   $n \ge 1$ (see Fig.~2 and Eq.~(3.3)) as:
  \begin{equation}
  {\cal M}^{(n+2)} = {\cal M}^{(0)}(L+V)^2L^{n-1}. 
   \end{equation}     
   Expressing ${\cal M}^{(n+2)}$ in terms of the gauge invariant quantity $B$
   and the gauge dependent quantity $L(\xi)$ gives:
\begin{equation}
  {\cal M}^{(n+2)} = {\cal M}^{(0)}\frac{1}{4}(B+L(\xi))^2L(\xi)^{n-1},
   \end{equation}
  which is manifestly gauge dependent.
  The Dyson sum in Eq.~(2.19) of Ref.~\cite{x24} correctly 
describes the all orders
  resummed amplitude, not for an arbitary gauge parameter $\xi$, but only for
  the special choice $\xi =-3$ when $V(\xi) = 0$, $B=L(-3)$ and
\begin{equation}
  {\cal M}^{(n+2)} = {\cal M}^{(0)}B^{n+1}~~~~~(\xi = -3) . 
   \end{equation}
   Clearly, the above argument for manifest gauge dependence, shown to be valid
   at the resummed one-loop level must also hold at arbitary loop order
   if the vertex and self-energy insertions satisfy a generalised Ward identity
   giving, at each order of perturbation theory, a condition such as (5.3).
   It has been shown~\cite{x32}, by the application of background
   field techniques, that Ward identities relating vertex and self-energy 
   contributions may indeed be derived that are valid to all orders in perturbation 
   theory. The gauge independence of the Ward identity at each fixed order 
   then necessarily implies gauge dependence when the corresponding vertex
   and self-energy diagrams are resummed. 
  \par The manifest gauge dependence of the 
     quark-quark scattering amplitude found, by direct calculation, in this
     paper, is, apparently, in contradiction with formal proofs~\cite{x49,x50}
      of the gauge
     invariance of $S$--matrix elements in non-abelian gauge theories.
     It seems however, that what is actually proved in these papers is
     the gauge invariance, at all orders in perturbation theory, of 
     generalised Ward identities. The consequences of {\it resumming} diagrams
     of fixed loop order, which as shown above, necessarily generates
     gauge dependence, were not considered.
For example, in the
     standard reference~\cite{x50}, the Lagrangian from
     which all the Feynman rules of the theory is derived is introduced, and the 
     change in this Lagrangian due to a change in the gauge parameter written down.
     It is then stated that the theory is gauge invariant if such a variation of
      the gauge parameter leaves $S$--matrix elements invariant. There immediately
      follows the statement: `We can formulate this condition' (i.e that the
       $S$--matrix elements are invariant) `in terms of a Ward identity that we have 
     written in terms of the diagrams in Fig.~2'. The unproved and unjustified
    assertion is thus made that the gauge invariance of a Ward identity is equivalent
    to gauge invariance of $S$--matrix elements. This will only be true
    of unresummed amplitudes at each loop order, not of the resummed amplitudes
    that, according to quantum mechanical superposition, must exist and are, indeed, essential
    to generate the RCC.
    In fact Ref.~\cite{x50} establishes {\it only} the gauge invariance of 
    Ward identities, and nothing else. As shown above, it is just the gauge invariance
    of the unresummed amplitudes that ensure the manifest gauge dependence of the
    resummed ones. Indeed an $S$--matrix element (even a formal, generic, one) appears
    nowhere among the equations of Ref.~\cite{x50}. In Ref.~\cite{x49} such
    a formal $S$--matrix element does appear, but its gauge invariance properties are 
    derived directly from a Ward identity. No actual physical process, and no effect
    of resummation, is considered.
   \par By consideration of a sub-set of $n$-loop diagrams for the
   off-shell gluon-gluon scattering amplitude in a non-covariant gauge, it has been
   claimed~\cite{x14} to demonstrate that the RCC of QCD is both gauge invariant
   and process independent, and that it may be identified with the 
solution (2.10) of the
   RGE (2.8). 
The $n$-loop diagrams considered are those that may be constructed as
   a formal `product' of $n+1$ tree level four-point functions. The diagrams contain
   both resummed one-loop gluon vacuum polarisation and vertex diagrams and a sub
   set of irreducible $n$-loop diagrams. This set of diagrams cannot, as claimed, 
  be identified with the one-loop RCC in Eq.~(2.10), which results
   solely from the resummation of one-loop (one particle irreducible) diagrams.
   At any order in the perturbation series these resummed one-loop diagrams
   contribute the leading powers of both $\ln Q$ and $n_f$. They give, 
in fact,
 the `renormalon' contribution~\cite{x26} (see above) that dominates the
   high order behaviour of the perturbation series. The irreducible $n$-loop
   ($n>1$) diagrams of the sub-set considered in Ref~\cite{x14} will contribute
   constant terms or non-leading powers of $\ln Q$, and therefore cannot be identified
   with terms in the diagrammatic expansion of the RCC in (2.10).
   Similarly, it has been conjectured (without explicit calculation) in
    Ref.~\cite{x31} that the `missing' vertex contributions needed to make, say,
    ${\cal M}^{(2)}$ in Eq.~(5.5) above, gauge invariant may be derived from `pinch
    parts' of two-loop irreducible diagrams. This is not possible since the 
    required `missing' contributions contain the factor $(\alpha_s \ln Q)^2$, (see 
    Eq.~(3.5), whereas, as is well known, in both QED~\cite{x33} and QCD~\cite{x34}
     irreducible two-loop vacuum polarisation and vertex diagrams have, at
    most, next-to-leading logarithmic behaviour $\approx \alpha_s^2 \ln Q$. 
    This is easily demonstrated by considering the two-loop solution 
    of the renormalisation group equation for the effective charge.
    In QED, or in QCD in the $\xi = -3$ gauge, Eq.~(4.2) generalises to:
    \begin{equation}
a_Q = \frac{a_{\mu}}{1+a_{\mu}\beta_0\lambda+\frac{\beta_1 a_{\mu}}{\beta_0}
     \ln(1+a_{\mu}\beta_0\lambda)}~,
    \end{equation} 
    where $\beta_1$ is the second $\beta$-function coefficient. Expanding the 
    right side of Eq.~(5.7) up to $O(a_{\mu}^3)$ yields:
 \begin{equation}
a_Q = a_{\mu} \left[ 1- a_{\mu}\beta_0\lambda+a_{\mu}^2\beta_0^2\lambda^2
- a_{\mu}^2\beta_1\lambda-a_{\mu}^3\beta_0^3\lambda^3 
+\frac{3}{2}a_{\mu}^3\beta_0^3\beta_1\lambda^2+ O(a_{\mu}^4)\right].
    \end{equation}
  It can be seen that $\beta_1$, given by two particle irreducible
  vacuum polarisation or vertex diagrams, occurs only in sub-leading
  logarithmic terms of the form $\beta_1 a_{\mu}^n \lambda^{n-1}$.
  No possible re-arrangement of these terms can compensate the manifest
  gauge dependence of the leading-logarithmic terms of the form
  $(\beta_0 a_{\mu} \lambda)^n$.                  
   \par The property exhibited above, for QCD, of gauge dependence of amplitudes
   on resumming one-loop corrections that, at lowest order, are gauge invariant, 
   is expected to be a general property of non-abelian gauge theories.
   In such theories the gauge boson propagator, in an arbitary covariant gauge
   is written as~\cite{x35}:
    \begin{equation}
 P^{\mu \nu}(q^2) = - \frac{i}{q^2-M^2}\left[ g^{\mu \nu}-(1-\xi)
 \frac{q^{\mu}q^{\nu}}{q^2-\xi M^2} \right] ,
 \end{equation}
 where $M$ is the renormalised gauge boson mass.
    The topographical structure
     of diagrams contributing to, say, neutrino-neutrino
    scattering via $Z$ exchange is the same as that for quark-quark scattering
    shown in Fig.~1. The one-loop vertex correction containing the non-abelian
    ${Z}{W}^+{ W}^-$ coupling is gauge dependent~\cite{x36}.
     Since the gauge dependence cancels at lowest order (without resummation)
    then, just as for QCD, it cannot cancel at any higher order in the resummed
    one-loop amplitude.
    Indeed a similar conclusion as to the necessity of the $\xi = -3$ gauge
    in order to obtain an effective charge that satisfies a RGE, reached in this
    paper for QCD, has previously been obtained, for the case of the Weinberg-Salam
     model, by Baulieu and Coqueraux~\cite{x37}. These authors pointed out
    that, with the special gauge choice (in the notation of the present paper)
    $\xi = -3$, the renormalisation constant of the ${Z}-\gamma$ mixing term
    vanishes, so that, in this case, effective charges satisfying separate
    (decoupled) RGE's
    may be associated the one-loop resummed photon and $Z$-boson propagators.
    It is also interesting to note that $\xi = -3$ is the unique choice of
    covariant gauge for which the photon mass counterterms in the 
    renormalised Lagrangian vanish.
    The case of $W$ exchange was not considered, but (as may be seen by
    inspection of the relevant formulae given, in an arbitary covariant gauge,
    in Ref.~\cite{x38}) the choice $\xi = -3$ results in the vanishing of the
    renormalisation constants associated with the one-loop vertex corrections
    to both the $Z$ and $W$ exchange fermion-fermion scattering amplitudes.
    As for the QCD case considered in the present paper, it is then expected
    that, only for this special choice of gauge, an effective charge satisfying
    a RGE may be associated with the resummed $W$ propagator. 
     \par The pinch technique, and 
    similar methods to formally shift gauge dependent pieces between diagrams,
    have also been applied to 
    electroweak amplitudes~[40,46-49]. Although gauge invariant
    boson proper self energy functions may be defined at one-loop level,
    any resummed higher order amplitude is demonstrably gauge dependent,
    by the same argument as that given above for QCD. So, although for a 
    particular choice of gauge parameter (such that the sum of all vertex
    corrections vanish) resummed $W$ and $Z$ running propagators may be
    defined that satisfy a RGE, they cannot, contrary to the claim
    of Refs.~[40,46-49], be so defined in a gauge invariant manner.          
    The diagrammatic inconsistency of these procedures 
     is made manifest by the inclusion in the modified vector
     boson self energy function of box diagram contributions. If the effective
     charge is expanded as a perturbation series, the box diagram contributions
     at each order will give terms of a geometric series. There is no way that 
     such a series can be meaningfully interpreted in terms of
     a sum of such diagrams required by quantum mechanical superposition. In fact, the contributions
     to physical amplitudes of box 
     diagrams can be systematically resummed~\cite{x42,x43,x44}, but the 
     result found is typically the exponential of a double logarithm
     of the relevant physical scale, not the sum of a geometric series. For the 
     case of the fermion-fermion scattering amplitude the contribution
     of box diagrams is expected to be important only in the $|t| \rightarrow
     0$ limit and to vanish~\cite{x38} in the $|t| \rightarrow \infty$ limit.
     \par A discussion of the gauge dependence, beyond one-loop order, of the 
     resonant $Z$ boson amplitude, may be found in Ref.~\cite{x45}. 
    \par The limited convergence domain imposed by requiring finiteness of the
    geometric series, which occurs in all theories in which the RCC decreases
    with increasing scales, can be avoided by choosing a very high renormalisation 
    scale\footnote{I am indebted to W.~Beenakker for this remark.}.
    This is equivalent, for such theories, to the choice, in QED, of on-shell
    renormalisation, yielding a RCC that is convergent for all scales below
    the Landau scale~\cite{x12}. Although such a choice guarantees convergence
    for all physical scales below the the chosen renormalisation point, it appears
    artificial from a physical viewpoint. If a
    strong interaction process at, say, the scale of the
    mass of the charm quark is to be described using a renormalisation point at the
    GUT scale $Q_{\rm GUT}$, the formula for the RCC at scale $m_c$ will depend upon
    the masses of all strongly interacting elementary particles
    below $Q_{\rm GUT}$ and  above $m_c$. There will be a phenomenon of `inverse decoupling'
    whereby the lower the scale the more high mass particles must be taken into
    account. Feynman amplitudes using such a renormalisation scale would loose
    their corresponence (valid in the on-shell scheme) with space-time processes.
    With on-shell renormalisation, the decoupling of heavy particles at low
    scales is understood in terms of a natural hierachy of physical scales. It seems 
    reasonable that the physics of the strong interaction
    at the scale of the charm quark mass, $m_c$, should be independent of the value
    of the top quark mass, $m_t$, when
    $m_t \gg m_c$. According to the the Uncertainty Principle, the contribution to
    vacuum polarisation loops of particles with masses much greater than the 
    propagator virtuality are expected to correspond to short lifetime 
    fluctuations giving only a small contribution to the radiative correction.
    This is no longer the case if the RCC is renormalised at scales
    $ \gg m_t$.
    \par An enormously successful phenomenology of the strong interaction
    has been developed over the past three decades based on 
    perturbative QCD. Many aspects of QCD, such as: the existence of the colour quantum
     number, of spin one gluons with self-coupling, the predicted values of
     colour factors, and a coupling constant that decreases with increasing
     scales in the experimentally accessible region, are confirmed, beyond doubt,
     experimentally~\cite{x46}. 
    However, it might be hoped that the physical predictions of a candidate 
    gauge {\it theory} would be gauge invariant at all orders in
    perturbation theory, as is the case in QED.  Explicit calculation for
    the current
    non-abelian gauge theories (both QCD and electro-weak theory), as
    reviewed in the present paper, seems to show,
    however, that this is not the case. The point with error bars at
     $Q = 90$~GeV on the loop gauge curve in Fig.~3 shows the uncertainty on 
     $\alpha_s$ ($\pm0.005$) of an early measurement using
     hadronic $Z$ decays at LEP~\cite{x47}.
     The input value $\alpha_s(5~$GeV$) = 0.2$ has been chosen to be consistent
     with deep inelastic scattering measurements~\cite{x48}. If the
     asymptotic running coupling
     constant with 
     time-like argument measured at LEP at the scale $Q=M_{Z}$ has
     a similar value to  $\alpha_s^{eff}(Q)$ for space-like
     argument considered in this paper, it is clear from Fig.~3 that the {\it measured}
     value of the gauge parameter must be close to $-3$. 
   
     \par On a more positive note, the phenomenological success, in QCD,
      of calculations based on the renormalisation group, as well as of `renormalon'
      based models, both of which are shown here to require the use of the $\xi = -3$ gauge, suggests
      that further understanding requires a deeper theoretical explanation of
      nature's apparent choice of this gauge (see Fig.~3). The question is, why, in
     non-abelian theories, do UV divergent loop diagrams apparently acquire logarithmic 
     corrections after renormalisation but not similar vertex diagrams? 
     Since the non-abelian triple gluon vertex occurs in both loop
     (vacuum polarisation) and vertex insertions it seems that consistency with
     experiment requires the cancellation of vertex contributions when summed over the
     complete particle content of the theory. That is, in some conjectured modified 
     version of QCD, complete cancellation of triangle vertex insertions similar to the
     anomaly cancellation provided by the particle content of the fermion families
    of the Standard Model.
      However, consistency
     with the diagrammatic description (i.e. with quantum mechanical superposition) must 
    always limit the scale-range of applicability of the RCC due to the convergence
     properties of geometric series.
    \par Finally, two important caveats concerning the work presented in this paper should be
      mentioned. Firstly, only covariant gauges specified by the parameter  $\xi$ are considered,
     whereas certain non-covariant gauges~\cite{JO} such as axial gauge ($A_3 = 0$) or light-cone gauge
     ($n_{\mu}A^{\mu} = 0$, $n^2 = 0$) are freqently employed in QCD phenomenology. The second concerns
      the recently developed successful application of Analytical Perturbation Theory (APT) to QCD 
      phenomenology~\cite{SS}.
      In order to avoid infra-red divergences when $\Lambda \rightarrow 0$ in the 
        conventional QCD perturbation series for the RCC,  APT introduces, by way of the 
         K\"{a}llen-Lehmann spectral representation of the dressed gluon propagator, the condition
        of analyticity in the $Q^2$ variable. This also imposes a causality requirement.
          In this case, the correspondence between the so-obtained `Euclidean running couplant',
           $\alpha_E$, and the sum of a QCD perturbation series, in which the terms represent the 
           contribution of specific Feynman diagrams, breaks down. No statements may then be made
          concerning the gauge dependence and convergence properties of $\alpha_E$.
     \section{Acknowledgements}

   I wish to thank, especially, W.Beenakker for his careful reading of an early version
     of this paper 
     and for constructively critical comments. Discussions with M.Consoli are also 
     gratefully acknowledged. I thank M.Veltman for pointing out to me the work 
     contained in Ref.~\cite{x50}, and an anonymous referee for introducing me to
     the related work of Mikhailov discussed in Section 5.

\section{Appendix}
\renewcommand{\theequation}{A.\arabic{equation}}
\setcounter{equation}{0}

  \par The sum of the first $n$ terms of a geometric series with a negative
   common ratio $r = -|r|$ is given by the relation~\cite{x15}:
\begin{equation}
 S_n = 1-|r|+|r|^2-...+(-|r|)^{n-1} = \frac{1-(-|r|)^{n}}{1+|r|} .
\end{equation}
 It follows that:
\begin{equation}
 S_n = \frac{1+|r|^{n}}{1+|r|} ~~~~~(n-{\rm odd}) ,
\end{equation}
\begin{equation}
 S_n = \frac{1-|r|^{n}}{1+|r|} ~~~~~(n-{\rm even}) .
\end{equation}
  If $|r|<1$ , than $S_{\infty}\equiv  ({\rm limit~as}~n \rightarrow \infty$ of $ S_n$)
   = $1/(1+|r|)$. If $|r|=1$ then $ S_n= 1$ for $n$ odd and  $ S_n= 0$ for $n$ even.
   If $|r|>1$,  $S_{\infty} = +\infty$ for $n$ odd and $S_{\infty} = -\infty$ for $n$ even.
   \par Table 4 presents values of $S_n$ versus $n$, demonstrating the convergence of the series
   for $|r|=1/2$  and its divergence for $|r|=2$. The Dyson sum of vacuum polarisation 
   insertions in the gluon propagator in QCD gives a geometric series similar to (A.1) above.
   The impossiblity of `asymptotic freedom', which conjectures that $S_{n} \rightarrow 0$
  as $n \rightarrow \infty$ is evident from inspection of (A.2) and (A.3) above and Table 4. 
    
\pagebreak

\newpage  
{\bf \Large Tables}\\

\begin{table}[h]
\begin{center}
\begin{tabular}{|c|c|c|c|} \hline
   i  &  Diagram in Fig. 1 & Type & $(\alpha_s^{\mu}/\pi)^{-1} \times$ Correction factor, $A_i$, 
   to ${\cal M}^{(0)}$  \\ \hline
   1 & b) + 13 $\leftrightarrow$ 24 & abelian vertex & 
  $ \frac{C_A}{2} \ln \frac{Q}{\mu}$ \\ 
   2 & c) + 13 $\leftrightarrow$ 24 & non-abelian vertex & 
    $-\frac{3 C_A}{2} \ln \frac{Q}{\mu}$    \\ 
   3 & d) & quark loops & 
     $\frac{n_f}{3} \ln \frac{Q}{\mu}$   \\
   4 & e) + f) + g) & gluon and ghost loops & 
    $-\frac{5 C_A}{6} \ln \frac{Q}{\mu}$ \\
\hline
\end{tabular}
\caption{UV divergent (before renormalisation) one-loop virtual corrections, 
at leading logarithmic accuracy,
to the $90^{\circ}$ quark-quark scattering amplitude
 in Feynman gauge. }      
\end{center}
\end{table}
\begin{table}[h]
\begin{center}
\begin{tabular}{|c|c|c|c|c|} \hline
Gauge & Feynman ($\xi=1$) & 
 Landau ($\xi=0$)  & Loop ($\xi=-3$)  & Vertex ($\xi=19/9$)  \\ \hline
$Q_L$(GeV) & 7.68$\times10^8$ & 1.02$\times10^5$  & 301 & $\infty$  \\
\hline
\end{tabular}
\caption{ Values of the Landau scale $Q_L$ (convergence limit) of the QCD effective charge
for $\mu = 5$ GeV, $\alpha_s^{\mu}= 0.2$, $n_f = 5$ }      
\end{center}
\end{table}
\begin{table}[h]
\begin{center}
\begin{tabular}{|c|c|c|c|c|} \hline
Gauge & Feynman ($\xi=1$) & Landau ($\xi=0$)  & Loop ($\xi=-3$)  & Vertex ($\xi=19/9$)  \\ \hline
$Q_0$(TeV) & 176 &  Undefined & Undefined &  18.1  \\
\hline
\end{tabular}
\caption[]{ Values of the $Q_0$ (the scale at which $\alpha_s^{eff}(Q)$ 
vanishes ). Parameters as in Table~2.}      
\end{center}
\end{table}

\begin{table}
\begin{center}
\begin{tabular}{|c|ccccc|} \hline
{ }&\multicolumn{2}{|c|}{$\gamma_{(1)}(2)~(NLO)$}&\multicolumn{3}{c|}{$\gamma_{(2)}(2)~(NNLO)$} \\ \hline
 Gauge & \multicolumn{1}{|c|}{$C_FC_A$}&\multicolumn{1}{c|}{$C_FN_f$}&\multicolumn{1}{c|}{$C_FC_A^2$}&
 \multicolumn{1}{c|}{$C_FC_AN_f$}&\multicolumn{1}{c|}{$C_FN_f^2$} \\ \hline
  & & & & &  \\
$\xi = 1$ & 13.9 & -$\frac{64}{27}$ & 117.70 & -38.50 & -$\frac{224}{243}$ \\
 Ref.~\cite{LRV} & & & & & \\ 
  & & & & &  \\
  $\xi = -3$ & 11.3 & -- & -76.0 & 23.0 & -- \\
 Ref.~\cite{Mikhailov}& & & & &  \\ 
  & & & & &  \\
$\xi = 0$ & 7.6 & -- & -13.2 & 12.4 & -- \\
Ref.~\cite{Mikhailov}& & & & &   \\
\hline
\end{tabular}
\caption[]{Contributions to the $n = 2$ moment of the non-singlet anomalous
    dimension in deep inelastic scattering on a nucleon due to 
  different classes of Feynman diagrams and for different choices of covariant gauge, $\xi$. See
  text for discussion.}      
\end{center}
\end{table}

\begin{table}[h]
\begin{center}
\begin{tabular}{|c||c|c|c|c|c|c|c||c|}\hline
$n$ & 2 & 3 & 4 & 5 & 6 & 7 & 9 & $1/(1+|r|)$\\
\hline \hline
 $|r| = 1/2$ & 0.5 &0.75 & 0.625 & 0.688 & 0.656 & 0.673 & 0.664 & 0.667\\
 $|r| = 2$ & -1 & 3 & -5 & 11 & -21 & 43 & -85 & 0.333\\
\hline
\end{tabular}
\caption{ Values of the sum, $S_n$, of first $n$ terms of the 
  geometric series in (A1) for $|r| = 1/2$ (convergent series) 
and  $|r| = 2$ (divergent series).}      
\end{center}
\end{table}

 \newpage
{\bf \Large Figure Captions}\\
{\bf Fig. 1~~~} Diagrams contributing to ${\cal M}^{(LO)}$. Solid lines denote quarks,
 wavy lines gluons and the closed loop in g) ghosts.
\newline
\newline
{\bf Fig. 2~~~} The topographical structure of diagrams contributing to the resummed
 quark-quark 
scattering amplitude: a) O($\alpha_s^2$), b) O($\alpha_s^3$), c) O($\alpha_s^4$).
In diagrams containing only one vertex insertion the contribution given by the
exchange $13 \leftrightarrow 24$ (see Fig. 1) is understood to be included. 
\newline
\newline
{\bf Fig. 3~~~} The variation of the Effective Charge (3.5) with the scale $Q$
for different choices of the gauge parameter $\xi$. 5 GeV $< Q <$ 300 GeV
 ($\alpha_s^{eff}(5~\rm{GeV}) = 0.2$, $n_f = 5$). The error bars ($\pm$ 0.005) on the 
 point at $Q = 90$ GeV on the loop gauge curve are typical of those on an
 $\alpha_s$ measurement using hadronic Z decays~\cite{x47}.    
\newline
\newline
{\bf Fig. 4~~~}  The topographical structure of diagrams contributing to: a) the
  O($\alpha_s^3$)
contribution to the anomalous magnetic moment of a quark, b) the four-loop
photon proper self energy function.
\newline
{\bf Fig.~5~~~} The topographical structure of diagrams contributing to non-singlet anomalous
  dimensions in deep inelastic scattering: a) leading order, b) next-to-leading order, c) and d)
  next-to-next-to-leading order. See text for discussion.

\begin{figure}[htbp]
\begin{center}
\mbox{\epsfig{file=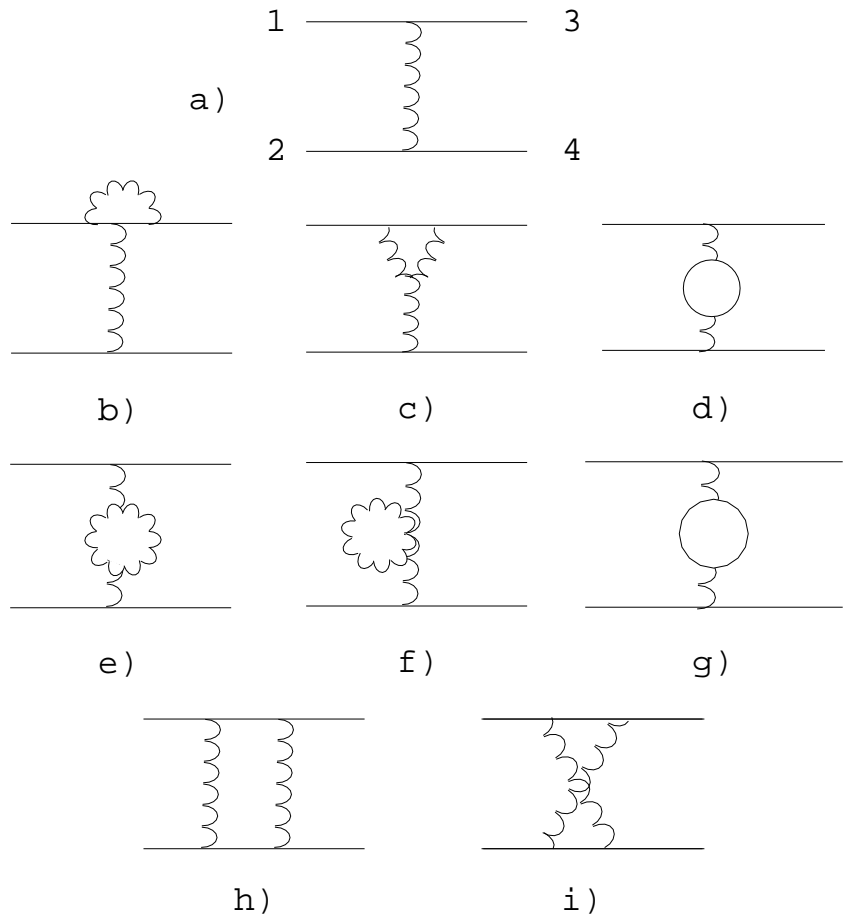,height=12cm}}
\caption{ Diagrams contributing to ${\cal M}^{(LO)}$. Solid lines denote quarks,
 wavy lines gluons and the closed loop in g) ghosts.}
\label{fig-fig1}
\end{center}
 \end{figure}  
\begin{figure}[htbp]
\begin{center}\hspace*{-0.5cm}\mbox{
\epsfysize10.0cm\epsffile{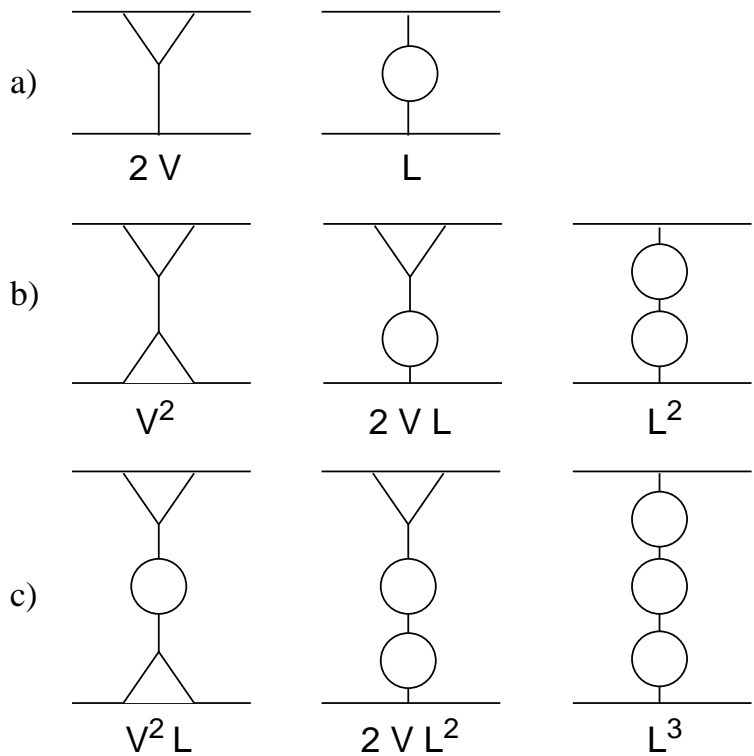}}
\caption{The topographical structure of diagrams contributing to the resummed
 quark-quark 
scattering amplitude: a) O($\alpha_s^2$), b) O($\alpha_s^3$), c) O($\alpha_s^4$).
In diagrams containing only one vertex insertion the contribution given by the
exchange $13 \leftrightarrow 24$ (see Fig. 1) is understood to be included.}
\label{fig-fig2}
\end{center}
 \end{figure}  
\begin{figure}[htbp]
\begin{center}\hspace*{-0.5cm}\mbox{
\epsfysize10.0cm\epsffile{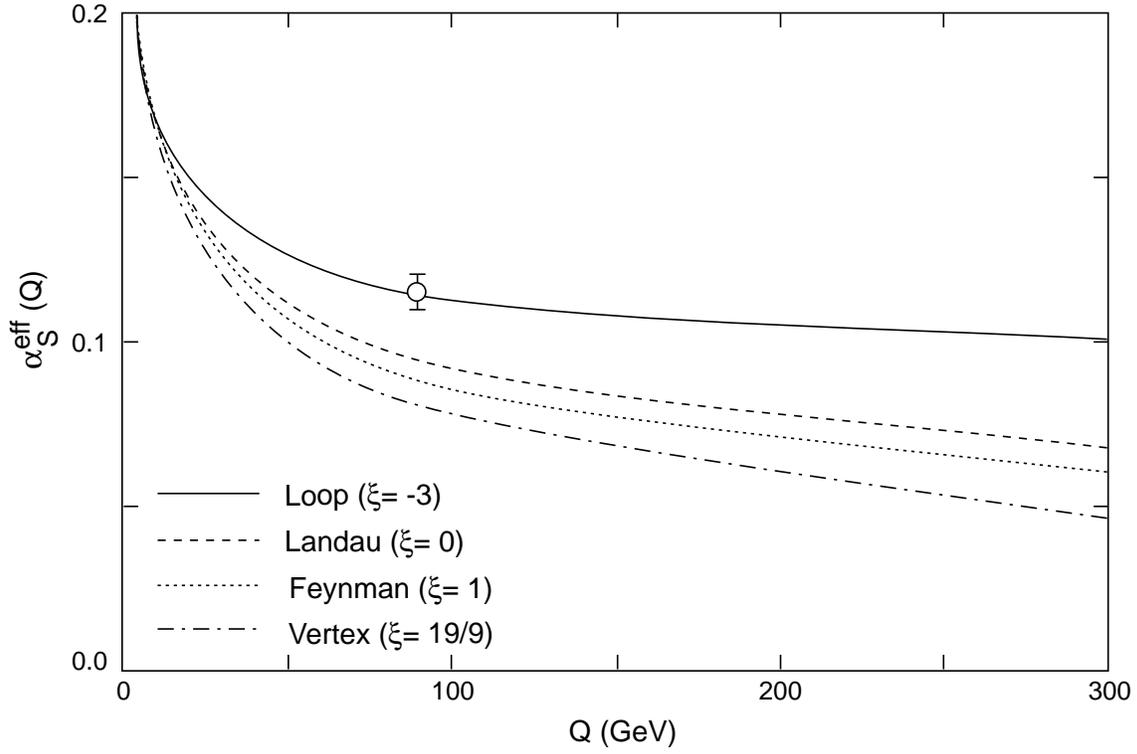}}
\caption{The variation of the Effective Charge (3.5) with the scale $Q$
for different choices of the gauge parameter $\xi$. 5 GeV $< Q <$ 300 GeV
 ($\alpha_s^{eff}(5~\rm{GeV}) = 0.2$, $n_f = 5$). The error bars ($\pm$ 0.005) on the 
 point at $Q = 90$ GeV on the loop gauge curve are typical of those on an
 $\alpha_s$ measurement using hadronic Z decays~\cite{x47}.}
\label{fig-fig3}
\end{center}
 \end{figure}  

\begin{figure}[htbp]
\begin{center}\hspace*{-0.5cm}\mbox{
\epsfysize10.0cm\epsffile{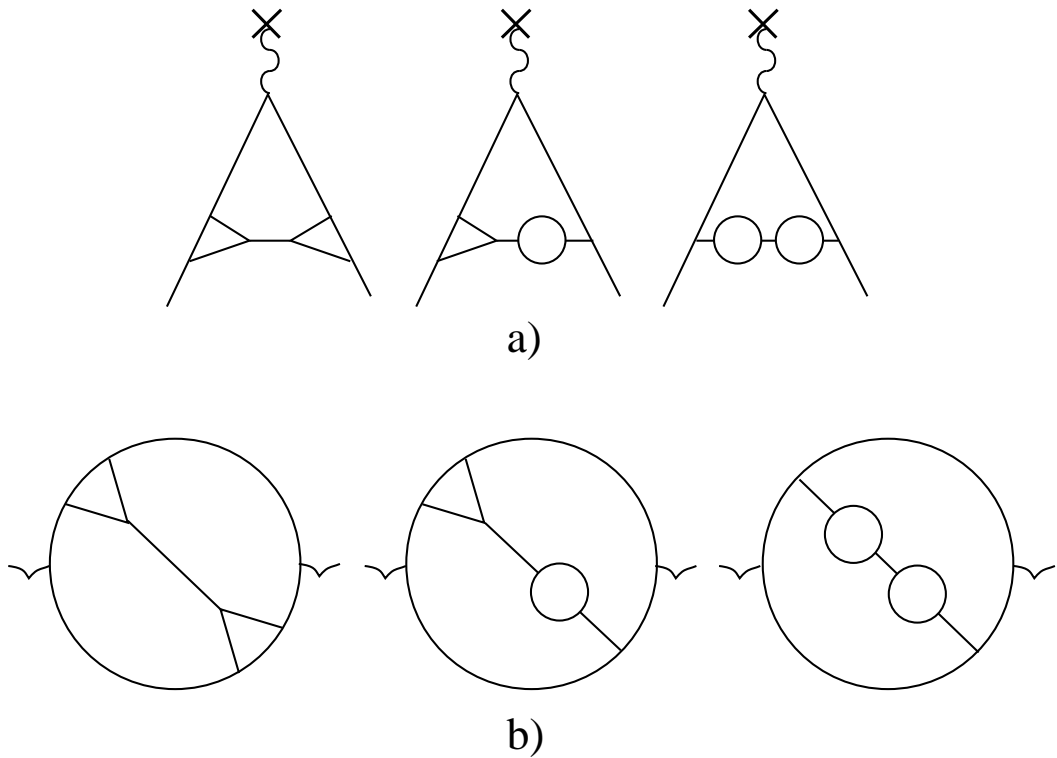}}
\caption{ The topographical structure of diagrams contributing to: a) the
  O($\alpha_s^3$)
contribution to the anomalous magnetic moment of a quark, b) the four--loop
photon proper self energy function.}
\label{fig-fig4}
\end{center}
 \end{figure} 

\begin{figure}[htbp]
\begin{center}\hspace*{-0.5cm}\mbox{
\epsfysize15.0cm\epsffile{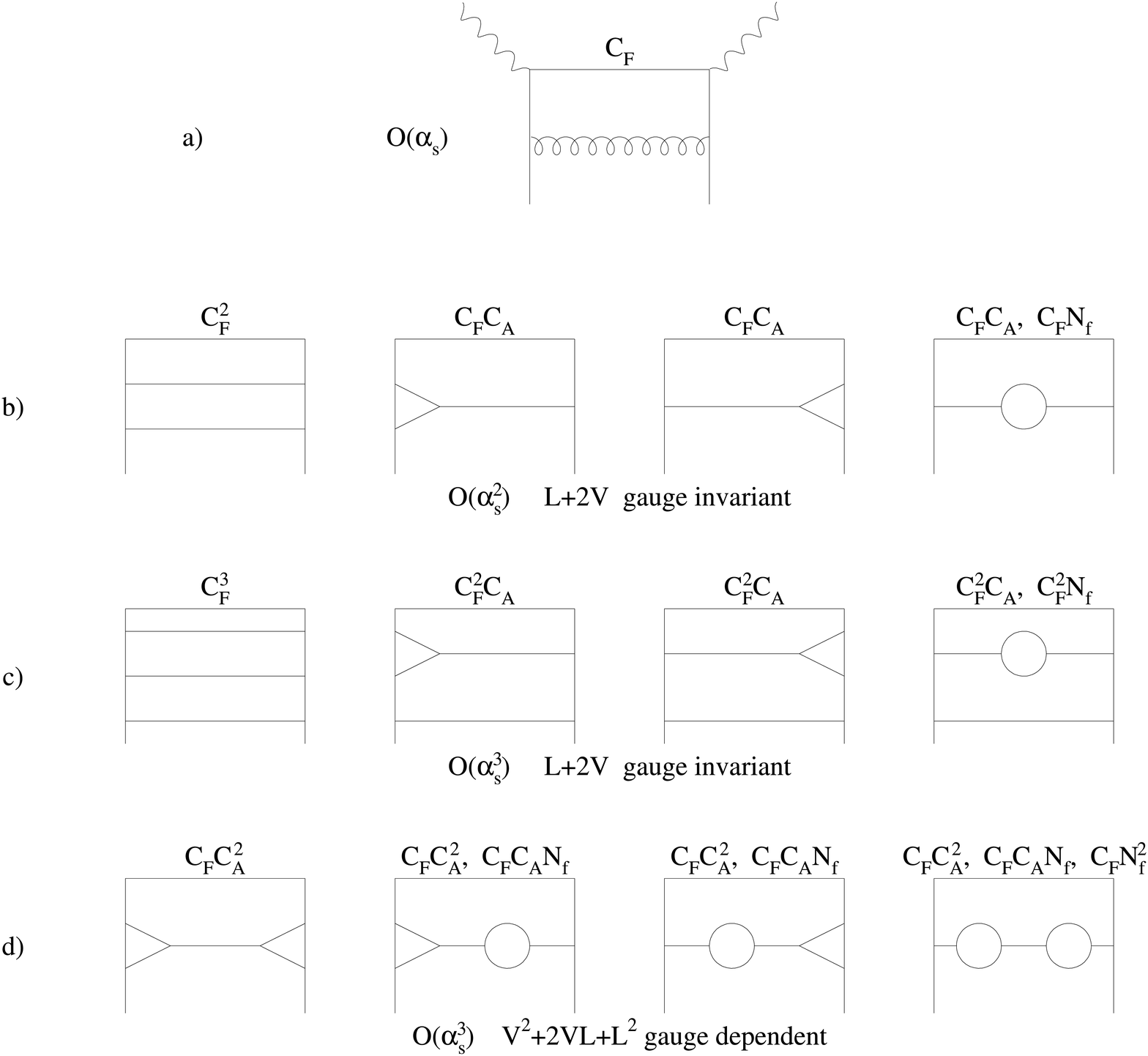}}
\caption{ The topographical structure of diagrams contributing to non-singlet anomalous
  dimensions in deep inelastic scattering: a) leading order, b) next-to-leading order, c) and d)
  next-to-next-to-leading order. See text for discussion.}
\label{fig-fig5}
\end{center}
 \end{figure}

\end{document}